\documentclass[preprint2]{aastex}
\usepackage{natbib}
\usepackage{epsfig}
\usepackage{graphicx}
\input epsf.sty

\shorttitle{3-D Accretion flows}
\shortauthors{Janiuk et al.}

\begin{document}

\title{Time Evolution of the 3-D Accretion Flows: Effects of the Adiabatic Index and Outer Boundary Condition}

\author{Agnieszka Janiuk\altaffilmark{1}, Maciej Sznajder\altaffilmark{1,2}, 
Monika Mo\'scibrodzka\altaffilmark{2} and  Daniel Proga \altaffilmark{3}}
\altaffiltext{1} {Copernicus Astronomical Center, Bartycka 18, 00-716 Warsaw, Poland}
\altaffiltext{2} {J. Kepler Institute of Astronomy, University of Zielona Gora, Poland}
\altaffiltext{3} {Department of Physics, University of Illinois, Urbana, IL, USA}
\altaffiltext{4} {Department of Physics and Astronomy, University of Nevada, Las Vegas, NV, USA}

\begin{abstract}
We study a slightly rotating accretion 
flow onto a black hole, using the fully three dimensional (3-D) 
numerical simulations.  We consider 
hydrodynamics of an inviscid flow, assuming a spherically symmetric 
density distribution at the outer boundary and a small, latitude-dependent 
angular momentum. We investigate the role
of the adiabatic index and gas temperature, and the 
flow behaviour due to non-axisymmetric effects.
Our 3-D simulations confirm 
axisymmetric results: the material that has too much angular momentum to be 
accreted forms a thick torus near the equator and the mass accretion rate is lower than the Bondi rate. 
 In our previous study 
of the 3-D accretion flows, for $\gamma=5/3$, 
we found that the inner torus precessed, even for axisymmetric conditions
at large radii. 
The present study shows that the inner torus precesses
 also for other values of the adiabatic index:
$\gamma=4/3, 1.2$, and 1.01.
However, the time for the precession to set increases with decreasing $\gamma$. 
In particular, for
$\gamma=1.01$ we find that depending on the outer boundary conditions, the 
torus may shrink substantially due to the strong inflow of the non-rotating matter
and the precession will have insufficient time to develop.
On the other hand, if the torus is supplied by the continuous inflow of the rotating material from the outer radii, its inner parts will eventually tilt and precess, as it was for the larger $\gamma$'s.
\end{abstract}


\section{Introduction}
\label{sec:intro}

The black holes capturing the gas from their vicinity are ubiquitous in the Universe,
and accretion onto the central black hole is the source of energy that powers 
the brightest sources on our sky, from quasars, through the gamma ray bursts, 
to the X-ray binaries. The fast rotation of material before it becomes 
captured by the central object leads to the formation of a thin accretion disk, 
where the gravitational energy is effectively dissipated through 
the viscous stresses and radiated away in the form of high energy radiation. 
On the other hand, if the gas does not rotate, or rotates very slowly, the 
viscous disk does not form and the material falls radially onto the center. The rate 
of accretion is then equal to the Bondi accretion rate, 
derived analytically (Bondi 1952). The flows with a small, latitude dependent
angular momentum were studied numerically
e.g. in Proga \& Begelman (2003a,b) in 2D, and Janiuk et al. (2008) in 3D, and showed 
that in this case the accretion rate 
is on the order of 20-30\% of the Bondi rate.
The astrophysical situation that may be described by such solutions is the 
radiatively inefficient flow of gas captured by the supermassive black holes in low 
luminosity AGNs, or the quiescent state of an X-ray binary.

The simulations show, that if the torus is formed due to rotation, 
the gas cannot accrete unless some angular momentum transport mechanism is introduced.
Therefore no accretion proceeds close to and through the equatorial plane, 
and the material accumulated in this region either 
flows out due to the centrifugal forces and gas pressure, or eventually 
turns to smaller latitudes and flows radially through the polar region.
The recent three dimensional studies showed, that in the latter case the amount 
of gas turning in the direction above or below the torus may depend on the 
azimuth, and in consequence lead to the tilt of this torus and its precession.

The present work is an extension of the previous study (Janiuk et al. 2008) to check how the gas 
properties, described by its equation of state, will influence the results.
The general form of the EOS assumed here is the polytropic relation, and the index 
of $\gamma = 5/3$, appropriate for gas pressure dominated system, is well 
suited for example in the inactive AGNs. On the other hand, the GRBs, 
which result from accretion of the massive stellar envelope, being a mixture of 
nucleons, neutrinos, electron-positron pairs and trapped photons 
(e.g. Janiuk et al. 2004; 2007), will rather be better described by a relativistic 
EOS with $\gamma=4/3$, than the ideal gas EOS. The still smaller values of $\gamma$, 
discussed in Mo\'scibrodzka \& Proga (2008), can account for the nearly isothermal 
gas, suggested to constitute the protogalactic disks.

Therefore we consider here the 3-dimensional hydrodynamical 
models with a range of $\gamma$ indices describing different gas microphysics.
We concentrate on the initially axisymmetric conditions, which were shown to produce
the non-axisymmetric results in the process of time evolution.
The content of the article is as follows. In Section \ref{sec:method},
we describe the method used in our calculations, 
to determine the initial conditions and 
subsequent evolution of the flow. 
In Section \ref{sec:results}, we present the simulations results 
for the 3-D models with various adiabatic indices (Sec. \ref{sec:indices}). 
We discuss the torus precession and the influence of the outer boundary conditions (Sec. \ref{sec:precess}), 
which occured to be important, especially for models with smaller $\gamma$.
The models with various gas temperatures are presented in Sec. \ref{sec:rbrs}.
We discuss our results in Section \ref{sec:diss}.

\section{Method}
\label{sec:method}

\subsection{Initial conditions}

In the analytical formula of Bondi (1952), the steady state solution
to the equations of mass and angular momentum conservation is parameterized
with the two quantities at infinity:
density, $\rho_{\infty}$, and sound speed, $c_{\rm s, \infty}$. The solution 
 is given by the accretion rate of:
\begin{equation}
\dot M_{\rm B} = \lambda 4 \pi R_{\rm B}^{2}\rho_{\infty} c_{\rm s, \infty}
\end{equation}
where the dimensionless parameter $\lambda$ depends only on the adiabatic index:
\begin{equation}
\lambda = {1 \over 4}{2 \over (5-3\gamma)}^{(5-3\gamma) \over 2(\gamma-1)}
\label{eq:lambda1}
\end{equation}
for $1<\gamma<5/3$.
In the limit of $\gamma \rightarrow 1.0$, this parameter approaches:
\begin{equation}
\lambda = {1\over 4} \exp(3/2),
\label{eq:lambda2}
\end{equation}
while in the limit of $\gamma =5/3$ the accretion rate is calculated from the 
transonic solution in the pseudonewtonian potential, PW, (Paczy\'nski \& Wiita 1980),
and depends on the sound speed at infinity (see Proga \& Begelman 2003a for the 
derivation from the density and sound speed at the sonic radius):
\begin{equation}
\lambda = {9\over 4} [{c_{\rm s, \infty} \over c} (1 + {1 \over \sqrt{3}}{c \over c_{\rm s, \infty}})]^{4}.
\label{eq:lambda3}
\end{equation}

The Bondi radius is equal to
\begin{equation}
R_{\rm B} = {G M \over c_{\rm s, \infty}^{2}}.
\end{equation}

The transonic solution is obtained for $\gamma < 5/3$ in the Newtonian potential, 
and the sonic radius is located at:
\begin{equation}
r_{\rm s} = {GM \over 2 c_{\rm s}^{2}(r_{\rm s})}
\end{equation}
with the sound speed in the sonic radius equal to:
\begin{equation}
c_{\rm s}(r_{\rm s}) = c_{\rm s, \infty} \sqrt {2 \over 5-3\gamma}.
\end{equation}
The above relation gives no sonic point for $\gamma=5/3$, and in this case the 
transonic solution is found for the PW potential, with the sonic radius equal to:
\begin{equation}
r_{\rm s} = {GM \over c^{2}} + \sqrt{2 \over 3} {GM \over  c_{\rm s, \infty} c}.
\end{equation}

We model the initial conditions for the spherically symmetric Bondi inflow,
solving iteratively the Bernoulli equation in 1-D:

\begin{equation}
B = H + {v_{\rm r}^{2} \over 2} + \Phi
\end{equation}
where 
\begin{equation}
\Phi = - {GM \over r - R_{\rm S}}
\end{equation}
is the PW potential, $R_{\rm S} = 2GM/c^{2}$ is
the Schwarzschild radius, and 
\begin{equation}
H = {1 \over \gamma-1} c_{\rm s, \infty}^{2} [({\rho \over \rho_{\infty}})^{\gamma-1} - 1]
\end{equation}
is the enthalpy.

We parameterize our model with  $\rho_{\infty}$ and  $c_{\rm s, \infty}$. The latter 
is conveniently used to define a dimensionless radius, 
$R_{\rm S}^{'} = R_{\rm S}/R_{\rm B} = 2 c_{\rm s, \infty}^{2}/c^{2}$.
Depending on $\gamma$, we 
calculate the dimensionless accretion rate from Eq. \ref{eq:lambda1}, 
Eq. \ref{eq:lambda2}, or Eq. \ref{eq:lambda3}. Then, we 
solve for the density profile, 
 while the velocity 
is found from the continuity 
equation:
\begin{equation}
\dot M_{\rm B} = -4 \pi r^{2} v_{\rm r}\rho.
\end{equation}

In the initial condition, we also impose a slight rotation of the outermost 
parts of the Bondi flow. This is done by means of the latitude-dependent angular 
momentum,
\begin{equation}
l = l_{0}(1-|\cos \theta|)
\end{equation}
where $l_{0}$ is a dimensionless parameter, which corresponds to the 
circularization radius $R_{\rm C}$ 
(i.e. the radius at which $GM/r^{2} = v_{\phi}^{2}/r$).
Therefore, our initial  velocity  $v_{\theta}$ is
everywhere equal to 0.0,  whereas the velocity $v_{\phi}$
is initially non-zero only in a fixed, quasi-conical region, with its
radial extension limited by the sound speed at infinity. 
This region is  defined as follows:
\begin{equation}
v_{\phi} = \left\{ \begin{array}{ll}
                 0 & {\rm for}~~\,~~|v_{\rm r}|>c_{\rm s, \infty} ~~\,~~\\  
     \sqrt{R_{C}} R_{B} c_{\rm s, \infty} {1-|\cos{\theta}| \over r
  \sin{\theta}}  &  {\rm for} ~~\,~~ |v_{\rm r}|<c_{\rm s, \infty}
\label{eq:vphi}
\end{array}
\right.
\end{equation}

\subsection{Boundary conditions}

The boundary conditions are set at the outer radius of the computational domain.
We impose a spherically symmetric outer boundary condition for the flow density, 
 $\rho(r_{\rm out})$, and energy density, $e(r_{\rm out})$,
 to be equal to these quantities determined from the initial 
analytical soulution for the Bondi flow. Because the outer radius of our grid is 
finite, the density at the outer edge
 is not exactly equal to $\rho_{\infty}$, but roughly twice larger. 
Nevertheless, the outer radius of 1.2 $R_{\rm B}$ is sufficient and such a condition
 matches well the outer, infinite and stationary sphere of gas, for which all the 
variables are determined analytically, with the inner, numerically simulated region. 

The boundary conditions for the velocity field are not specified otherwise than these 
incorporated in the ZEUS-MP code. We use the free inflow/outflow condition for the 
radial velocity, the reflection symmetry with respect to the polar axis, and periodic 
boundary condition in the azimuthal direction.

Note, that in this way in every time step we update the value of density at 
the outer radius, but we do not update the specific angular momentum.
In other words, the physical situation that we consider, is a spherically 
symmetric, stationary cloud of gas accreting from the infinity 
(where the density and temperature are given by $\rho_{\infty}$ and $c_{s,\infty}$) 
onto the central massive black hole. 
This cloud was at time $t=0$ perturbed by imposing some small amount of angular 
momentum, which made the cloud evolve.

\subsection{Time evolution of the flow}

In our calculations we use the 3-D code ZEUS-MP (Stone \&
Norman 1992; Hayes \& Norman 2003), modified by ourselves 
to incorporate the PW gravitational potential.
The computations were performed on the multi-processor computer cluster 
machines (see Catlett et al. 2007).
The ZEUS-MP code solves the equations of hydrodynamics:

\begin{equation}
{d\rho \over dt} + \rho \nabla {\bf v} = 0
\end{equation}
\begin{equation}
\rho {d {\bf v} \over dt} = -\nabla P +\rho \nabla \Phi
\end{equation}
\begin{equation}
\rho{d \over dt}({e \over \rho}) + P\nabla {\bf v} = 0
\end{equation}
where $\rho$ is the gas density,
 $e$ is the internal energy density, $P=(\gamma-1)e$ is the gas
pressure, and {\bf v} is the velocity of the flow.

The  other model parameters, $c_{s, \infty}$, $\rho_{\infty}$, $M_{\rm BH}$  are 
chosen such that the ratio of the Bondi radius to the Schwarzschild 
radius is fixed, and 
is equal to 1000, 300 or 100, depending on the model.
The radial grid is in the range from $r_{\rm in} = 1.5 R_{\rm S}$, to
$r_{\rm out} = 1.2 R_{\rm B}$.
The rotation of the flow is parameterized by the value of $R_{\rm C}$, and 
for most of the models it is equal to 0.1 (in the units of Bondi radius).

We use the spherical coordinate system, RTP, and
the resolution in $r$-direction was 140 zones, with $d r_{\rm i+1}/d
r_{\rm i} =1.05$, in $\theta$-direction it was 96 zones, and 
in $\phi$-direction it was 32 zones, with
$d \theta_{\rm j+1}/d \theta_{\rm j} = d \phi_{\rm k+1}/d
\phi_{\rm k} = 1.0$. The range of the grid in $\theta$ and $\phi$ directions is 
from 0 to $\pi$ and from 0 to 2$\pi$, respectively.

\section{Results}
\label{sec:results}

The hydrodynamical computations of an axisymmetric, 2.5-D model of an accretion flow
with low angular momentum were presented in PB03, and continued in Mo\'scibrodzka \& 
Proga (2008).
In the 3-D analysis presented in Janiuk, Proga \& Kurosawa (2008), 
we recalculated their 2.5-D models for one chosen value of the adiabatic index,
$\gamma=5/3$, and the sound speed at infinity corresponding to the
Bondi radius equal to 1000 $R_{\rm S}$. In that paper, we
confirmed that the 3-D effects, such as the
nonaxisymmetric distribution of the angular momentum in the accreting fluid, 
play an important role for a resulting rate of mass accretion with respect to the Bondi
one. We also found that a non-axisymmetric torus is subject 
to the tilt and precession due to the acoustic instabilities in the innermost gas.
These instabilities appear when any small asymmetry in the angular momentum 
distribution arises during the torus evolution. In the dynamical timescale, 
they lead to the torus misplacement from the equatorial plane and its subsequent 
precession.
We showed that the instability develops in a highly supersonic flow, whereas for small 
Mach numbers it is suppressed. We attributed this behaviour with the 
Papaloizou \& Pringle (1995) type of instability, which for low order modes is 
driven by the Kelvin-Helmholtz mechanism.

In all of our models, 
the simulations start from a spherically symmetric 
gas cloud around a black hole, with  density and velocity
distributions derived from the Bondi solution.
The matter located far from the black hole contains specific
angular momentum that exceeds the critical value, 
$l_{\rm crit}=2 R_{\rm S} c$, at the equatorial plane
and is decreasing towards the polar regions (cf. Eq. \ref{eq:vphi}).

The time evolution of the system with such initial conditions proceeds first through 
a short phase of the purely radial Bondi accretion, which duration depends
on the assumed distance of the initially rotating gas from the central black hole.
Then, the evolution of the flow switches to a long-term phase of 
torus accretion. In this phase, 
the gas settled near the equatorial plane is 
supported against gravity by the gas pressure and rotation,
 and the rate of accretion on the black hole, 
$\dot M_{\rm in}$, decreases below $\sim 30\%$ of the Bondi rate.
The reason is that the material from the polar regions is
still accreting radially, while the torus material is mainly rotating, and 
is either outflowing, or accretes after turning slighlty towards one of the poles.

Most of our simulation runs lasted up to
 about $1.5 \times 10^{4}$ dynamical time at 
the inner radius, $t^{'}=t_{\rm dyn}(r_{\rm in}) = 2\pi/\Omega_{K}(r_{\rm in})$, 
where $\Omega_{\rm K}$ is the Keplerian velocity.

The axisymmetry of the solution 
 breaks after a certain time, depending on the model parameters (see below).
When this happens, the inner torus becomes tilted with respect to the
equator and it starts precessing. 
This may also be accompanied by the fluctuations of the accretion rate.

\subsection{Evolution of the flow for various adiabatic indices }
\label{sec:indices}

We performed the runs for the axisymmetric initial conditions 
for a range of $\gamma$  and gas temperature ($c_{\infty}$, reflected by the value of
$R_{B}/R_{S}$).
The models are summarized in Table 
\ref{tab:models}.

\begin{table}
\begin{center}
\begin{tabular}{l c c c c c r}     
\hline\hline
 Mo & $\gamma$ & $R_{B}/R_{S}$ & $T_{\rm end}$  & $\dot M$ & Prec  \\
\hline 
$A$ & 1.01& 1000  & $1.1\times 10^{4}$   & 0.12-1.55  & --  \\
$Av$& 1.01 & 1000 & $1.2\times 10^{4}$   & 0.1-0.75  & yes \\
$B$ & 1.2 & 1000 & $1.25\times 10^{4}$   & 0.09-0.61 & yes  \\
$C$ & 4/3 & 1000 & $1.25\times 10^{4}$   & 0.10-0.28 & yes \\
$D$ & 5/3 & 1000 & $1.25\times 10^{4}$   & 0.18-0.25 & yes \\
$E$ & 5/3 & 300   & $8.1\times 10^{3}$ & 0.30-0.42 & no   \\
$F$ & 5/3 & 100   & $6.2\times 10^{3}$ & 0.34-0.62 & no  \\
$G$ & 4/3 & 300   & $6.2\times 10^{3}$ & 0.12-0.32 & no   \\
$H$ & 4/3 & 100   & $6.1\times 10^{3}$ & 0.05-0.30 & no  \\
\hline
\end{tabular}
\end{center}
\label{tab:models}
\caption{Summary of the evolutionary models. The time is given in units of dynamical time at $r_{\rm in}$. The accretion rate $\dot M$ is given in the units of $\dot M_{\rm B}$.} 
\end{table}

 \begin{figure}
\includegraphics[width=7cm]{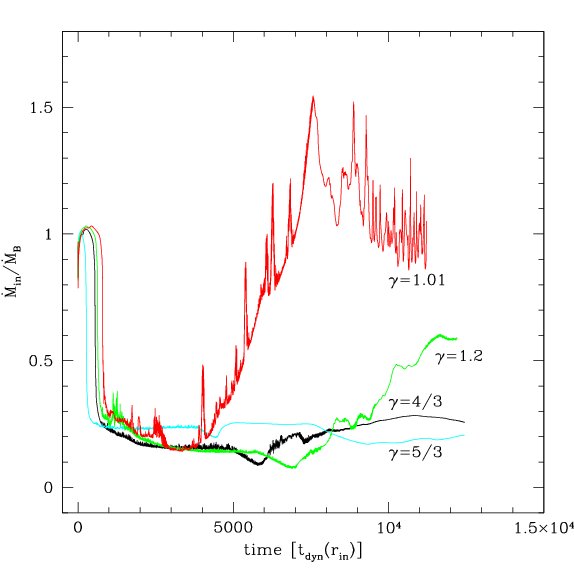}
\caption{Time evolution of the mass accretion rate
  through the inner
  boundary ($\dot M_{\rm in}$), in units of the
  Bondi rate ($\dot M_{\rm B}$). The models are: 
$A$ (red line),
$B$ (green line), $C$ (black line),
$D$ (cyan line).
}
  \label{fig:Mintime}
   \end{figure}


In Figure \ref{fig:Mintime} we show the time evolution of the
accretion rate through the inner boundary, $\dot M_{\rm in}$ 
(in units of the Bondi accretion rate), 
for various adiabatic indices. As the Figure shows, the accretion rate onto BH varies
in time. 
Before the rotating material approaches the black hole,
$\dot M_{\rm in}$
 is equal to the Bondi accretion
rate for all models. 
Once the gas starts rotating also in the innermost
parts, the accretion rate drops
to a small fraction of the Bondi rate. The moment when this happens, i.e. the end of the transient 
phase of the purely radial accretion,
depends on the adiabatic index, and is in the range from  about 
$t^{'}=3.2\times 10^{2}$ 
for $\gamma=5/3$,
up to about $t^{'}=9.5\times 10^{2}$ for $\gamma=1.01$.

In all the models, the accretion rate after the torus has formed, is on average  
low, and does not exceed 30\% of the Bondi rate.
The fluctuations of the accretion rate for $\gamma=1.2$, 4/3 and 5/3
are not significant:
they are on the order of $0.02 - 0.04 \dot M_{\rm B}$.
However at the end of the simulation, for $\gamma=1.2$ the accretion rate
approaches and exceeds $0.5 \dot M_{\rm B}$.
For the almost isothermal model ($\gamma=1.01$), this happens much earlier.
The small accretion rate is kept only for a very short
time, $t^{'} \sim 3.9 \times 10^{3}$, and after 
that $\dot M_{\rm in}$ starts rising, to reach and exceed the Bondi rate.
This is accompanied by very large fluctuations of the accretion rate, 
in a form of pronounced flares
with amplitude of $0.3 - 0.5 \dot M_{\rm B}$.

\begin{figure}
\includegraphics[width=7cm]{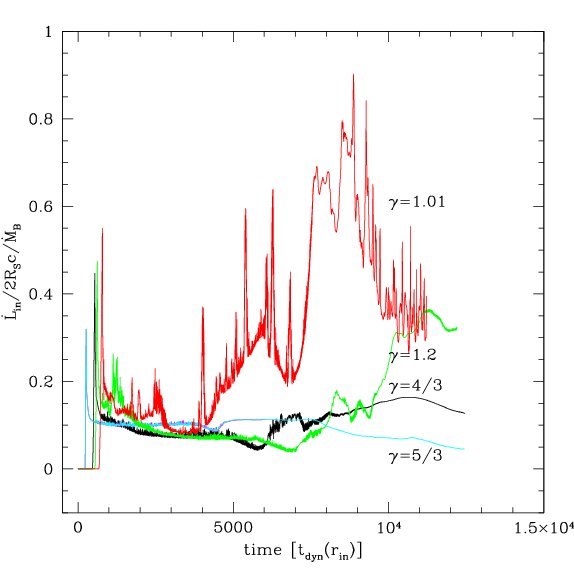}
\caption{Time evolution of the angular momentum flux through the inner
  boundary, in units of the critical angular momentum times the Bondi
  accretion rate. 
The models are: $A$ (red line),
$B$ (green line), $C$ (blackline),
$D$ (cyan line).
}
  \label{fig:Lintime}
   \end{figure}

In the Figure \ref{fig:Lintime}, 
we plot the evolution of the angular momentum flux through the inner
  boundary: $\dot L_{\rm in} = \int l_{\rm spec} \rho v_{r} ds$, in units of the
  critical angular momentum $l_{\rm crit}$ 
and renormalized by the
  value of the Bondi accretion rate.
At the beginning of the simulation, $\dot L_{\rm in} = 0$, because
in the vicinity of the black hole the gas is not rotating. Once the
  rotating matter reaches the inner boundary, the angular momentum
starts accreting to the center, and  $\dot L_{\rm in}$ sharply increases. 
  However, after several orbital cycles the outflow begins and the net radial
  velocity as well as the density near the polar regions drop, so
 $\dot L_{\rm in}$ decreases to a moderate value of $\sim 0.1$ (in the units of 
$l_{\rm crit} \dot M_{\rm B}$). This is the case for the
gas and ratiation pressure dominated models, $\gamma=5/3$ and $\gamma=4/3$, and 
also for the model with $\gamma=1.2$ for most of the simulation.
Only for the model with $\gamma=1.01$, the angular momentum flux rises after $t\sim 3.9\times 10^{3}$
and strongly oscillates.

Note that the quantity plotted in the Figure \ref{fig:Lintime} is
  a flux of total angular momentum, not specific angular momentum.
This corresponds to the amount of angular momentum which may be 
 transferred to the black hole and used to spin it up.
Figure \ref{fig:Lintime} can be used to calculate the total angular momentum  which the
  black hole could gain during our simulation. We find, after the proper 
unit conversions, that this number is extremely low:
$a = (cJ)/GM^{2} \approx 4\times10^{-6} - 10^{-5}$, where 
$J=\dot L_{\rm in} \Delta t$.
For the isothermal model, $\gamma=1.01$, the amount of the total angular 
momentum accreted onto the 
black hole will be 2-3 times larger (cf. Fig. \ref{fig:Mintime}).
Still, it does not lead to the spin-up of the black hole, as the material which
 is being accreted is the very slowly rotating gas mostly from the polar regions.
In the models with $\gamma = 5/3, 4/3$ and 1.2,
the accretion rate and $L_{\rm in}$ are only slightly  variable.

\begin{figure}
\epsscale{0.85}
\plotone{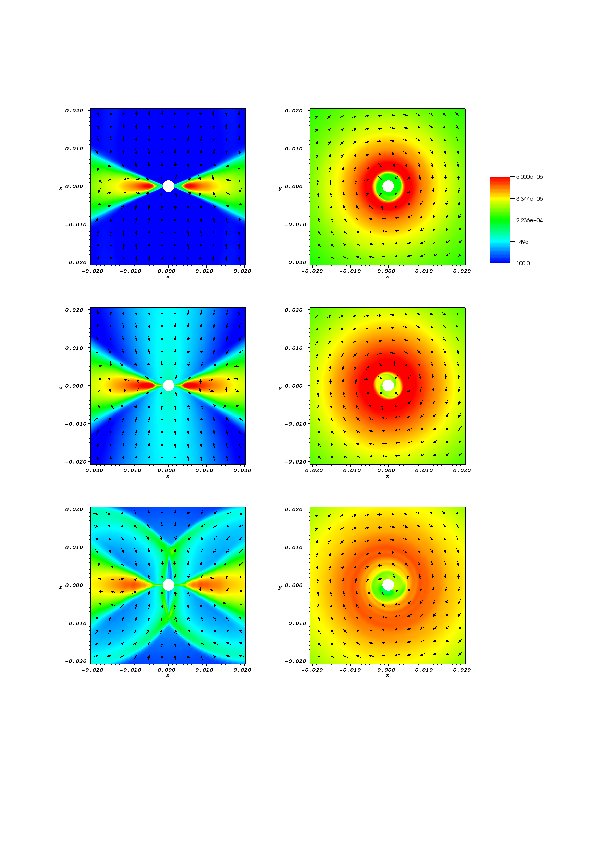}
\caption{Contour density maps and velocity field in the inner region 
for the model A ($\gamma=1.01$), in the side-view (along the x-axis; left)
and in the top view (along the z-axis; right).
The time snapshots are taken at $t^{'} = 3\times 10^{3}, 6.3\times 10^{3}$ and $1.1\times 10^{4}$, from top
to bottom. The units on the x, y and z axes are in $R_{\rm B}$.
}
  \label{fig:Maps_1x01}
   \end{figure}

\begin{figure}
\epsscale{0.85}
\plotone{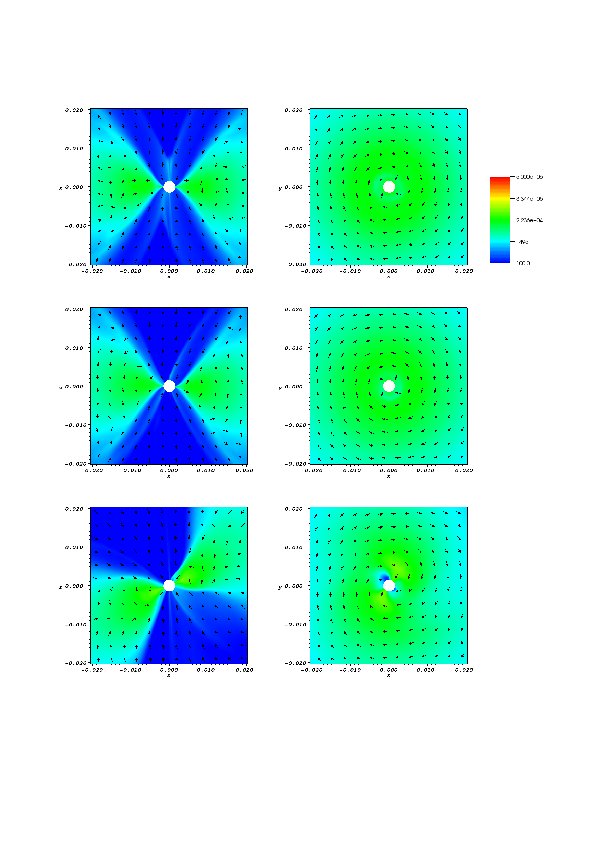}
\caption{Contour density maps and velocity field in the inner region 
for the model B ($\gamma=1.2$), in the side-view (along the x-axis; left)
and in the top view (along the z-axis; right).
The time snapshots are taken at $t^{'} = 3\times 10^{3}, 6.3\times 10^{3}$ and $1.1\times 10^{4}$, from top
to bottom. The units on the x, y and z axes are in $R_{\rm B}$.
}
  \label{fig:Maps_1x2}
   \end{figure}

\begin{figure}
\epsscale{0.85}
\plotone{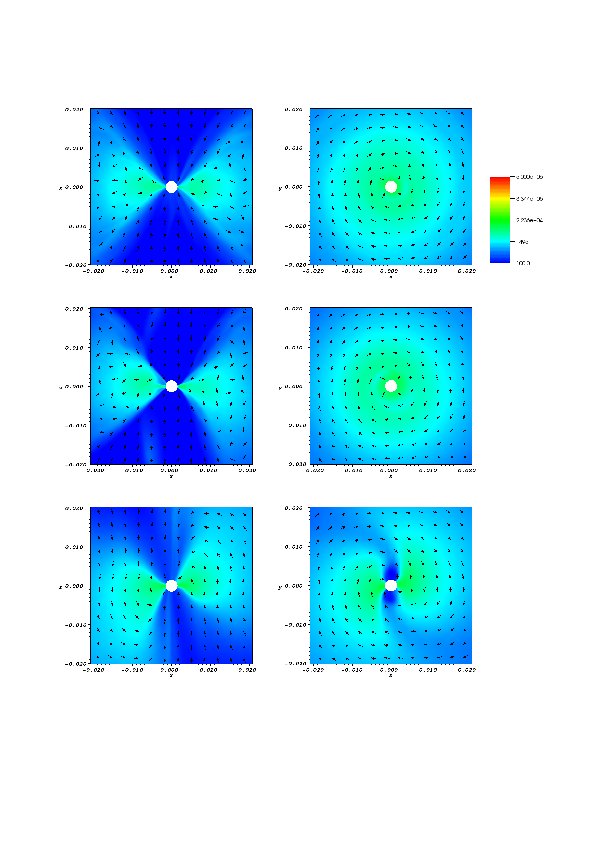}
\caption{Contour density maps and velocity field in the inner region 
for the model C ($\gamma=4/3$), in the side-view (along the x-axis; left)
and in the top view (along the z-axis; right).
The time snapshots are taken at $t^{'} = 3\times 10^{3}, 6.3\times 10^{3}$ and $1.1\times 10^{4}$, from top
to bottom. The units on the x, y and z axes are in $R_{\rm B}$.
}
  \label{fig:Maps_4x3}
   \end{figure}

\begin{figure}
\epsscale{0.85}
\plotone{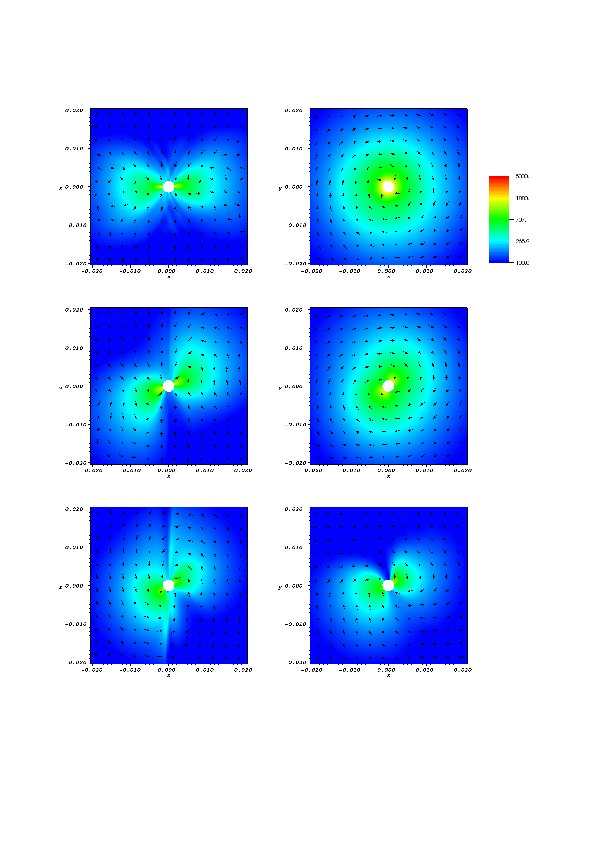}
\caption{Contour density maps and velocity field in the inner region 
for the model D ($\gamma=5/3$), in the side-view (along the x-axis; left)
and in the top view (along the z-axis; right).
The time snapshots are taken at $t^{'} = 3\times 10^{3}, 6.3\times 10^{3}$ and $1.1\times 10^{4}$, from top
to bottom. The units on the x, y and z axes are in $R_{\rm B}$.
}
  \label{fig:Maps_5x3}
   \end{figure}

To understand the behaviour of the accretion rate $\dot M_{\rm in}$, 
we need to study the structure
of the flow. In Figures \ref{fig:Maps_1x01}, \ref{fig:Maps_1x2}, \ref{fig:Maps_4x3} and
\ref{fig:Maps_5x3}, we plot the contour maps of the density and velocity fields
in the innermost regions of the flow (i.e. up to 20 $R_{\rm S}$), 
for the models A,B,C and D, respectively. The maps are plotted in three
time-snapshots, and in the two views: equatorial plane and side-on.

In Figure \ref{fig:Maps_1x01} we show the density contour maps for the 
inner region of the accretion flow, for the model A ($\gamma=1.01$).
The maps also show the velocity field, and are
 plotted  for several snapshots during
the evolution: $t^{'}\approx 3 \times 10^{3}$, $t^{'}\approx 6.3 \times 10^{3}$ and
$t^{'}\approx 1.1 \times 10^{4}$.
The panels on the left, show the side-on view of the inner torus, i.e. 
the $x-z$-plane, while the
panels on the right show the top-view, i.e. the equatorial plane.

The torus, which formed after the transient phase of the Bondi accretion,
is very dense: the maximum density in the equatorial plane exceeds 
$8\times 10^{6} \rho_{\infty}$.
At the time $t^{'}\sim 3\times 10^{3}$, the polar regions are clean, and most of the 
material is 
settled near the equator. The radial infall of material onto the central black hole 
goes mainly 
through the poles, while the torus material is turbulent and due to the centrifugal 
barrier the gas may flow 
outwards. At this time, the rate of accretion through the inner boundary, 
$\dot M_{\rm in}$, is the lowest and is below 20\% of the Bondi rate. The 
rapid variations of the 
accretion rate
are caused by the turbulences. Later, the variations
 are enhanced by the azimuthal 
perturbations, which grow in the density field in the innermost region of the torus 
(see the middle-right panel of the Figure \ref{fig:Maps_1x01}, 
for time $t^{'}\sim 6.3\times 10^{3}$).
As was shown in Figure \ref{fig:Mintime}, in the late phase of the evolution in 
model A, 
the accretion rate $\dot M_{\rm in}$ steeply rises, 
reaches the Bondi rate, and rapidly varies about this value.
However, the flow does not return completely to a spherically symmetric configuration
of the Bondi type. The inner torus still exists, 
but becomes much smaller and 
its density is lower. The maximum density in the equatorial plane is now about 
$\sim 2 \times 10^{6} \rho_{\infty}$, 
while the density in the polar regions becomes larger.

Now the polar material forms a kind of streams, in 
which the density is about 100 times larger in the end of the run than it was during 
the 
phase of 'clean poles'. This large density, in addition to the highly supersonic 
radial velocities 
(see Fig. \ref{fig:Mach_1x01} below), are the cause for the net accretion rate to be 
large again. The streams are formed by the material that has too large angular momentum to accrete radially, but flows laminarly over the torus and turns toward the poles.
The 'arcs' visible on the density map are the results of the obligue shocks, where the $v_{\theta}$ velocity component is changing.

We note, that such kind of behaviour, i.e. the drammatic rise of the accretion rate
 in the nearly isothermal mode, $\gamma=1.01$, was not found in the study by Mo\'scibrodzka \& Proga (2008), however, their 2-D runs were performed within twice 
longer timescale, with $T_{\rm end}\sim 4\times 10^{4}$. In their simulations, 
 a twice smaller rotation parameter was used, $l_{0}=0.05$, so that the  accretion 
rate $\dot M_{\rm in}$ in their models did not drop below $\sim 0.4 \dot M_{\rm B}$.
We tested that the difference in the rotation parameter is not crucial:
we made the test calculations with other values of $l_{0}$ and
qualitatively the results were similar.
Also, we checked that it is not the role of the 3-D effects, that make the nearly isothermal torus smaller and weaker after a short time in our simulation.
The same model was tested in the 2.5-D axisymmetric configuration with ZEUS-MP, 
and we found that
the mean accretion rate also increased to about $\dot M_{\rm B}$ after 
$t^{'} \sim 6\times 10^{3}$, with the flaring behaviour being even more pronounced than  in our 3-D case.

 What we found crucial for the behaviour of the flow at small $\gamma$, was the 
treatment of the outer boundary condition for the azimuthal velocity.
In their simulations with ZEUS 2-D, Mo\'scibrodzka \& Proga (2008) imposed the
update of specific angular momentum at the outer boundary in every time step. 
In our simulations, we imposed the rotation of the outer regions only in the initial 
condition, to mimic the transient cloud of gas, that 
contains some angular momentum, passing by the vicinity of the galaxy center.
When the cloud sinks into the Bondi sphere, the remaining gas at infinity 
no longer rotates.
The Bondi sphere is still modeled 
with the boundary condition for density, and matched with the analytical Bondi 
solution.
 
We note here, that despite the rise of the accretion rate to the value of the Bondi 
rate, the torus in the innermost parts of the flow does not completely disappear. 
The torus shrinks in size and is less dense, because of the massive inflow of the 
non-rotating material from the outer boundary. However, it still acts as an obstacle 
for the gas flowing radially onto the black hole, which results in the 
rapid fluctuations of the accretion rate at the inner boundary.

 In Figure \ref{fig:Mach_1x01} we show the contour plots of the total Mach number,
 $M_{\rm tot} = |\vec v| / c_{\rm s}$ (left) and the radial Mach number, 
$M_{\rm r} = v_{\rm r}/c_{\rm s}$ (right), for $\gamma=1.01$.
The color contours show the constant Mach number of 0.5, 1.0 and 1.5 (red, green
and blue, respectively),
while the grey colors show the three chosen constant denity contours, to mark 
the position of 
the inner torus.
In model $A$, which is nearly isothermal, most of the flow is highly supersonic.
Starting from the Bondi solution, the sonic radius is located at a very large distance, $r_{\rm s} = 
0.49 R_{\rm B}$, and the total Mach number is everywhere by definition equal to the
radial one. When the torus forms, it is supported by rotation, and the radial velocity in the torus is smaller than the sound speed. Only in the innermost radii, below $\sim 2.6\times 10^{-3} R_{\rm B} =
2.6 R_{\rm S}$, the radial Mach number is greater than 1.0. 
The sonic radius for the total Mach number is however  at the same place as before, 
and the net velocity
of the flow is supersonic. At the time $t^{'} \sim 6.3\times 10^{3}$ (middle panels of the Figure \ref{fig:Mach_1x01}), the sonic radius in the equatorial region shrinks.
The radial and the total Mach numbers are both greater than 1.0 close to the poles. The supersonic velocity components closer to the equatorial plane are mostly non-radial.
Later during the simulation, this trend reverses.
The sonic surface for $M_{\rm tot}$ closes again at $\sim 0.5 R_{\rm B}$, and it coincides with the sonic surface for  $M_{\rm r}$ (not shown in the bottom-right figure, because of the zooming-in). 
Close to the equatorial plane, 
the region with subsonic radial velocities shrinks several times, and at the 
outer edge of the torus, the radial velocity is equal to the sound speed.
What is also worth noticing in Fig. \ref{fig:Mach_1x01}, is that the actual 
size of the constant density contours, that has shrunken about 10 times in 
the last time snapshot. 

 \begin{figure}
\epsscale{0.8}
\plotone{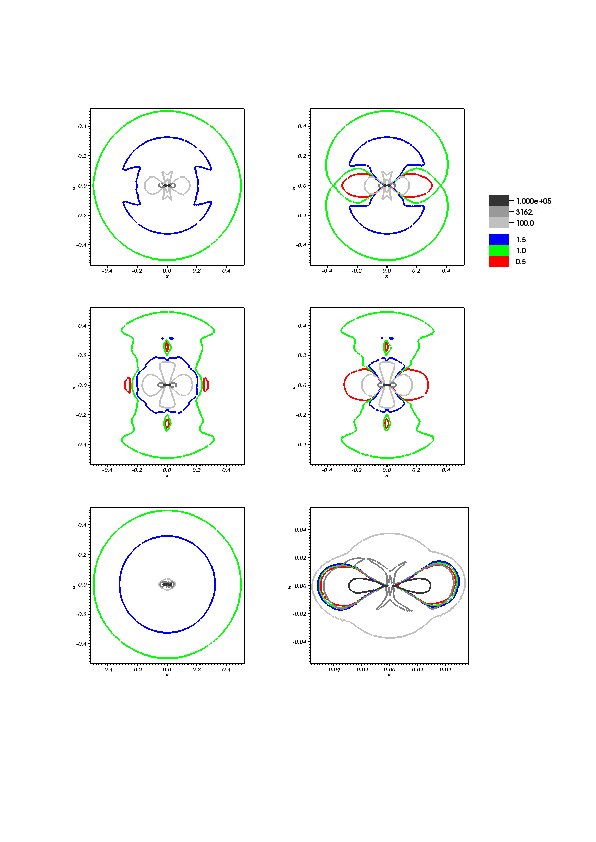}
\caption{Contour plots of the total Mach number, $M_{\rm tot}$ (left) and the radial Mach number, $M_{\rm r}$ 
(right) 
for the model A ($\gamma=1.01$).
The time snapshots are taken at $t^{'} = 3\times 10^{3}, 6.3\times 10^{3}$ and $1.1\times 10^{4}$, 
from top
to bottom. The units on the x and z axes are in $R_{\rm B}$, 
(the bottom-right panel is  zoomed-in).
The color contours show the constant Mach number of 0.5, 1.0 and 1.5 (red, green
and blue, respectively),
while the grey colors show the constant density contours, between $10^{2}$ and $10^{5} \rho_{\infty}$.
}
  \label{fig:Mach_1x01}
   \end{figure}

In Figure \ref{fig:Maps_1x2} we show the flow topology for the model B ($\gamma=1.2$).
In this model, the density of the torus is much lower. Initially, the flow is axially symmetric, but at the time $t' \sim 6.3\times 10^{3}$, the axial symmetry is broken. This is because the material at 
one side of the torus (i.e. the azimuth $\phi \sim 0^{\circ}$), tries to reach the black hole and flows in above the equatorial plane, while at the other side 
(i.e. the azimuth $\phi \sim 180^{\circ}$), the gas tries to flow in below the equatorial plane. This is shown by the directions of the arrows that represent the velocity field.
As was already shown in Janiuk et al. (2008), such a behaviour leads to the tilt of the torus, and here
the tilt is clearly visible at the time $t' \sim 1.1 \times 10^{3}$.

\begin{figure}
\epsscale{0.8}
\plotone{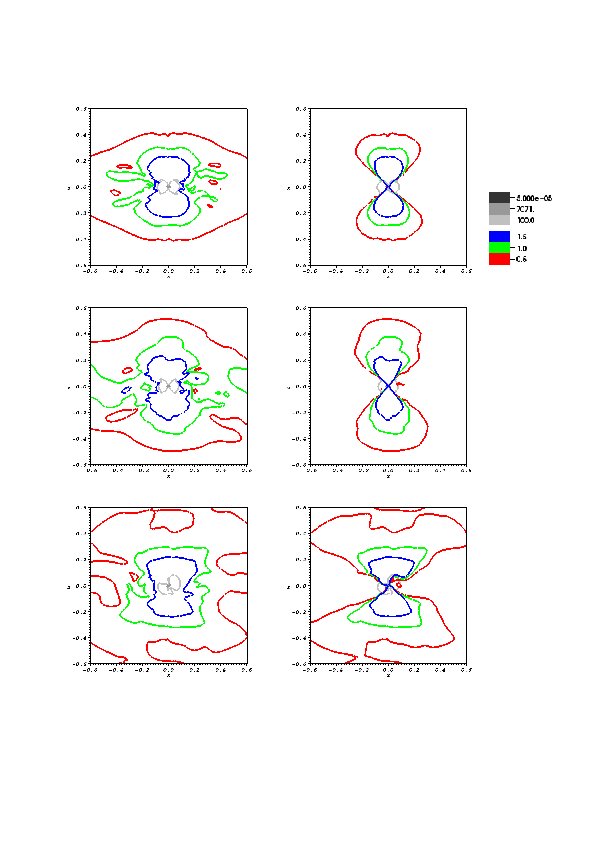}
\caption{Contour plots of the total Mach number, $M$ (left) and the radial Mach number, $M_{\rm r}$ 
(right) 
for the model B ($\gamma=1.2$).
The time snapshots are taken at $t^{'} = 3\times 10^{3}, 6.3\times 10^{3}$ and $1.1\times 10^{4}$, 
from top
to bottom. The units on the x and z axes are in $R_{\rm B}$.
The color contours show the constant Mach number of 0.5, 1.0 and 1.5 (red, green
and blue, respectively),
while the grey colors show the constant denity contours, between $10^{2}$ 
and $10^{5} \rho_{\infty}$.
}
  \label{fig:Mach_1x2}
   \end{figure}

In Figure \ref{fig:Mach_1x2} we show the Mach number contours for the model $B$, $\gamma=1.2$.
The color contours show the constant Mach number of 0.5, 1.0 and 1.5 (red, green
and blue, respectively),
while the grey colors show the constant denity contours, to mark the position of the inner torus.
The shape of the sonic surface is very different from the previous case (in model A), and resembles in shape the number ``eight''.
The radial velocity exceeds the speed of sound in the polar regions, while near the equator, somewhat irregular contour of the total Mach number shows the contribution from the supersonic, non-radial 
velocity components.
When the torus gets tilted, the innermost part of the sonic surface tilts as well, and the size of this surface slightly increases.

\begin{figure}
\epsscale{0.8}
\plotone{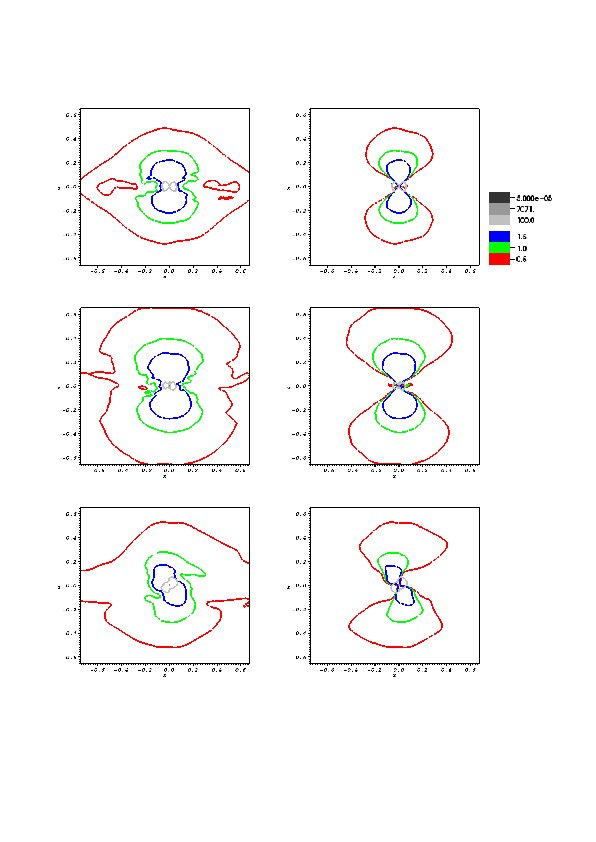}
\caption{Contour plots of the total Mach number, $M$ (left) and the radial Mach number, $M_{\rm r}$ 
(right) 
for the model C ($\gamma=4/3$).
The time snapshots are taken at $t^{'} = 3\times 10^{3}, 6.3\times 10^{3}$ and $1.1\times 10^{4}$, 
from top
to bottom. The units on the x and z axes are in $R_{\rm B}$.
The color contours show the constant Mach number of 0.5, 1.0 and 1.5 (red, green
and blue, respectively),
while the grey colors show the constant denity contours, between $10^{2}$ 
and $10^{5} \rho_{\infty}$.
}
  \label{fig:Mach_4x3}
   \end{figure}

\begin{figure}
\epsscale{0.8}
\plotone{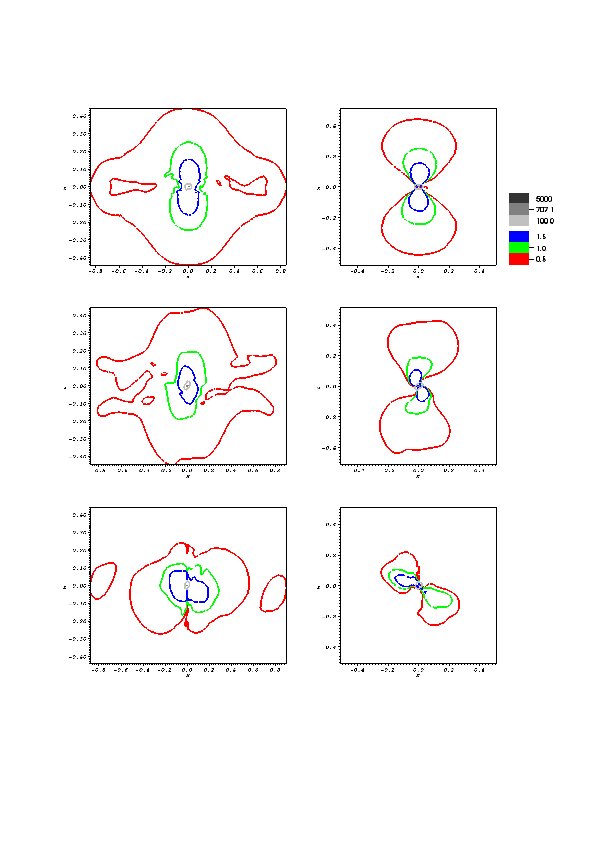}
\caption{Contour plots of the total Mach number, $M$ (left) and the radial Mach number, $M_{\rm r}$ 
(right) 
for the model D ($\gamma=5/3$).
The time snapshots are taken at $t^{'} = 3\times 10^{3}, 6.3\times 10^{3}$ and $1.1\times 10^{4}$, 
from top
to bottom. The units on the x and z axes are in $R_{\rm B}$.
The color contours show the constant Mach number of 0.5, 1.0 and 1.5 (red, green
and blue, respectively),
while the grey colors show the constant denity contours, between $10^{2}$ 
and $10^{5} \rho_{\infty}$.
}
  \label{fig:Mach_5x3}
   \end{figure}

Similar results are obtained for the adiabatic indices $\gamma=4/3$ and $\gamma=5/3$. 
The density maps are shown in Figures \ref{fig:Maps_4x3} and \ref{fig:Maps_5x3}, while
the Mach number contours are shown in Figures \ref{fig:Mach_4x3} and \ref{fig:Mach_5x3}.
We note that qualitatively, the basic pattern of the torus evolution
is uniform for $\gamma=1.2, 4/3$ and 5/3 in the time scale of our simulations.
We note that there is a systematic decrease in the torus densities with increasing 
$\gamma$ (in Fig. \ref{fig:Maps_5x3} the color scale had to be
 changed to make the density maps more contrasted). 

We also note here that in the late time evolution of model $D$, seen in Fig. 
\ref{fig:Maps_5x3} (bottom-left panel), the flow pattern is strongly affected by the
poles. This is the result of a discontinuity at the z-axis, characteristic for the
RTP coordinate system. The boundary conditions in the theta direction, which we 
use in our calculations (reflection with inversion in 2nd and 3rd velocity component),
are appropriate
 as long as the flow is basically axisymmetric. 
Because in the late 
time evolution the torus tilts, the z-axis is no longer the symmetry axis. 
Especially, for the model D the tilt is largest (see  below, Sec. \ref{sec:precess}), 
so the pole effects for this $\gamma=5/3$ are the strongest at the end of the evolution.

\begin{figure}
\epsscale{0.9}
\plotone{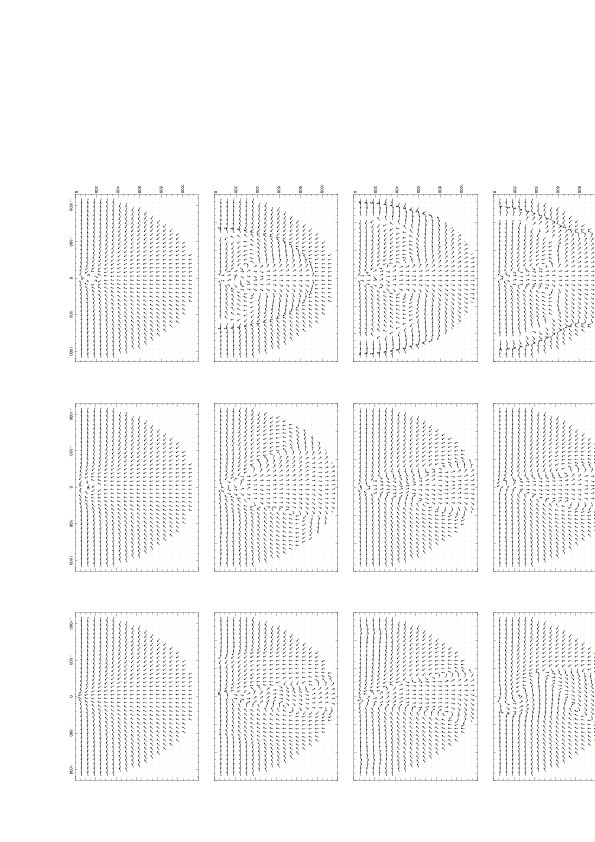}
\caption{Sequences of the poloidal velocity fields, taken for the azimuth $\phi=0$, 
for the models with $\gamma=1.01, 1.2, 4/3$ and 5/3 (from left to right) and at three different time snapshots:
$t^{'} = 3\times 10^{3}, 6.3\times 10^{3}$ and $1.1\times 10^{4}$ (from top to bottom).
}
  \label{fig:velfields}
   \end{figure}

To summarize and compare the evolution of the flow for various adiabatic indices,
we show the large scale poloidal velocity fields, plotted for models A,B,C and D, at three representative time snapshots.
Figure \ref{fig:velfields} shows, from top to bottom, the
velocity fields at time  
$t^{'} = 3\times 10^{3}, 6.3\times 10^{3}$ and $1.1\times 10^{4}$, taken at the azimuth
$\phi=0$. Because the model is not axially symmetric, the snapshots taken at different
azimuths will not look exactly the same. Nevertheless, the global pattern, which depends mainly on $\gamma$, is very similar.
The first column shows the model with $\gamma=1.01$. Here, the torus which formed initially in the equatorial region, is accompanied by the departure of the 
velocity field from the purely radial. The gas turns back due to the centrifugal force, however no net outflow occurs, and the largest size of the turbulent region is
 about 200 $R_{\rm Schw}$. In the last time snapshot, for $\gamma=1.01$ 
the torus is indeed very small, and the size of the turbulent region is only about 20 
$R_{\rm Schw}$, hardly visible in the scale of the Figure.

For other models, the turbulent region is much larger, and grows with $\gamma$.
The shock front between the outflowing and inflowing gas is clearly visible, and this front propagates outwards with time. The large outflow of material occurs from the equatorial region, and is accompanied by the large scale turbulences.

\subsection{Torus precession}
\label{sec:precess}

The precession of the torus is discussed below in
terms of the tilt and twist angles.
These angles are defined as (see e.g. Fragile et al. 2007; Janiuk et al. 2008):
\begin{equation}
\beta(r,t) = \arccos\Big({L_{z} \over L}\Big)
\label{eq:tilt}
\end{equation}
and
\begin{equation}
\gamma(r,t) = \arccos\Big({L_{x} \over \sqrt{L_{x}^{2}+L_{y}^{2}}}\Big).
\label{eq:twist}
\end{equation}

Initially, the flow is axisymmetric and the only component of the angular 
momentum vector is $L_{\rm z}$ 
(which sign depends on the direction of the flow rotation), while 
$L_{\rm x}$ and $L_{\rm y}$ are basically zero.
Therefore the tilt angle is vanishingly small and the disk does not
precess.
After some time, the non-zero $L_{\rm x}$ and $L_{\rm y}$ components
appear in the flow, while $|L_{\rm z}|$ decreases, and the rotation axis tilts towards 
the $x-y$ plane. Then, $L_{\rm y}$ and $L_{\rm x}$ are changing periodically,
with $L_{\rm z}$ being almost constant, i.e. the rotation axis 
moves clockwise or counter-clockwise, depending on the model.

 \begin{figure}
\plottwo{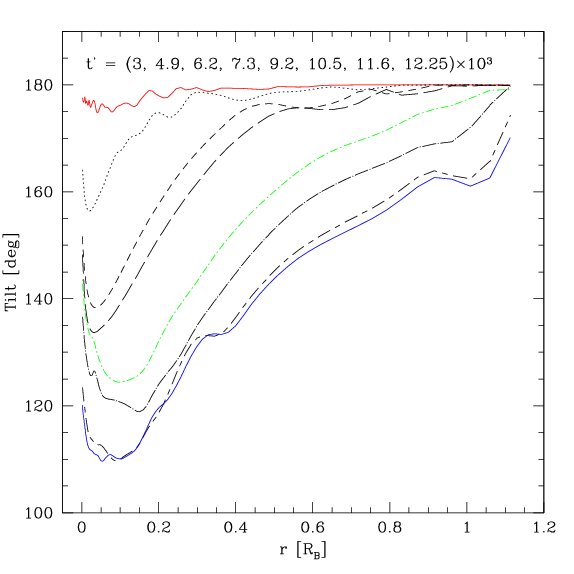}{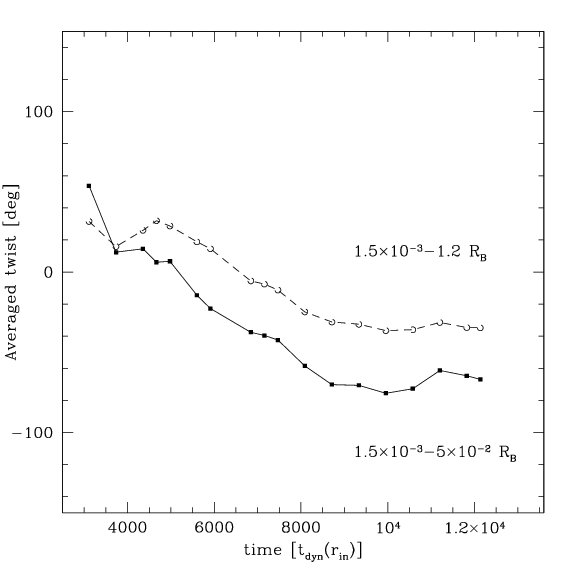}
\caption{The tilt (left panel) and averaged twist (right panel) angles,
calculated for model with $\gamma=5/3$. The tilt is shown as a function of radius,
for several time-snapshots, given in the top of the panel. The initial tilt is of 
180$^{\circ}$, meaning that the angular momentum vector is not tilted to the z-axis
 and the 
flow rotates with a negative azimuthal velocity.
The twist angle is shown as a function of time, and the two curves 
represent the averaging over radius from $r=r_{\rm in}$ to $r=0.05 R_{\rm B}$ 
(solid line) and to $r=1.2 R_{\rm B}$ (dashed line).
The twist angle decreasing from $\sim 60^{\circ}$
to  $\sim -80^{\circ}$ (innermost torus) and from
$\sim 40^{\circ}$ to  $\sim -40^{\circ}$ (total)
reflects the clockwise, initially differential precession.
}
  \label{fig:tilt_5x3}
   \end{figure}

 \begin{figure}
\plottwo{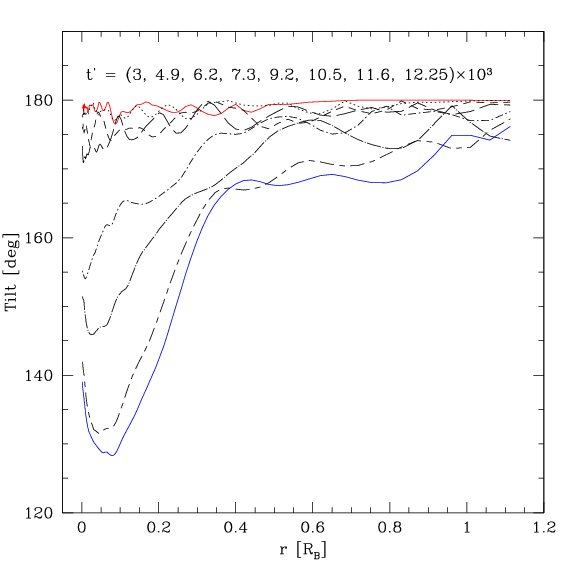}{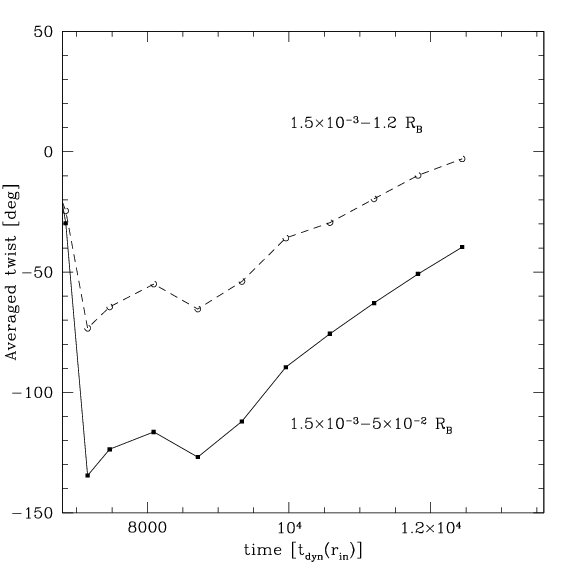}
\caption{The tilt (left panel) and averaged twist (right panel) angles,
calculated for the model
with $\gamma=4/3$. The tilt is shown as a function of radius,
for several time-snapshots, given in the top of the panel. The initial tilt is of 
180$^{\circ}$ means that the angular momentum vector is not tilted to the z-axis 
and the 
flow rotates with a negative azimuthal velocity.
The twist angle is shown as a function of time, and the two curves 
represent the averaging over radius from $r=r_{\rm in}$ to $r=0.05 R_{\rm B}$ 
(solid line) and to $r=1.2 R_{\rm B}$ (dashed line).
The twist angle increasing from $\sim -120^{\circ}$
to  $\sim -40^{\circ}$ (innermost torus) and from
$\sim -70^{\circ}$ to  $\sim 0^{\circ}$ (total)
reflects the counter-clockwise, initially differential precession.
}
  \label{fig:tilt_4x3}
   \end{figure}

\begin{figure}
\plottwo{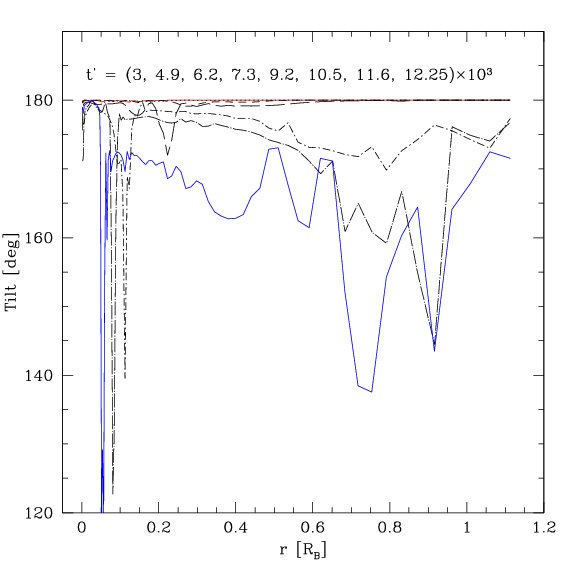}{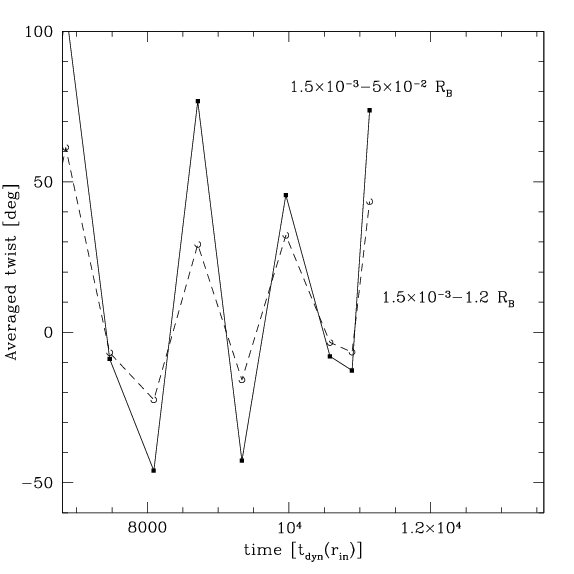}
\caption{The tilt (left panel) and averaged twist (right panel) angles,
calculated from the time evolution of the  model $A$
($\gamma=1.01$). The tilt is shown as a function of radius,
for several time-snapshots, given in the top of the panel. The initial tilt is of 
180$^{\circ}$ means that the angular momentum vector is not tilted and the 
flow rotates with a negative azimuthal velocity.
The twist angle is shown as a function of time, and the two curves 
represent the averaging over radius from $r=r_{\rm in}$ to $r=0.05 R_{\rm B}$ 
(solid line) and to $r=1.2 R_{\rm B}$ (dashed line).
The angular momentum vector is always almost perpendicular to the equatorial plane, and 
despite the large scatter in the two angles, the apparent small tilt
 is due to the  very small components of $L_{\rm x}$ and $L_{\rm y}$. 
}
  \label{fig:tilt_1x01}
   \end{figure}

The evolution of the tilt and twist angles proceeds as follows.
The tilt (i.e. the angle
between the angular momentum vector of the gas and the $z$ axis), is initially equal 
to zero for all radii. Later during the simulation the tilt rises strongly
in the inner parts of the flow, while it is negligible in the outer parts.
The twist angle is defined as a cumulative angle by which the angular momentum 
vector revolves in the $x-y$ plane by the time $t$.
Before the disk was tilted, it did not precess, and by definition 
the twist was zero everywhere. When the tilt increases, the twist angle rises
fast in the innermost parts of the flow, which we show in the Figure \ref{fig:tilt_5x3} (as well as in the following Figures \ref{fig:tilt_4x3}, \ref{fig:tilt_1x2} and 
\ref{fig:tilt_1x01}) as the twist
averaged over the radius, from $r_{\rm in}$ to $0.05 R_{\rm B}$. 
Initially, the strongest rise for the innermost radii implies the differential 
precession of the torus.
The outermost parts of the flow have a negligible precession, and 
above $\sim 0.5 R_{\rm B}$ the twist oscillates around zero. The total twist, i.e. 
the average over 
the radius from $r_{\rm in}$ to $r_{\rm out}$, is smaller than that for the innermost 
torus.

In the Figure \ref{fig:tilt_5x3} we show the tilt and twist evolution for the model with
$\gamma=5/3$. The tilt is shown as a function of radius,
for several time-snapshots. The initial tilt of 
180$^{\circ}$ means that the angular momentum vector is not tilted and the 
flow rotates with a negative azimuthal velocity.
The twist angle is shown as a function of time, and the 
averaging over radius was calculated in two ranges: 
from $r=r_{\rm in}$ to $r=0.05 R_{\rm B}$ 
(solid line) and to $r=1.2 R_{\rm B}$ (dashed line).
The twist angle decreasing from $\sim 60^{\circ}$
to  $\sim -80^{\circ}$ (innermost torus) as well as from
$\sim 40^{\circ}$ to  $\sim -40^{\circ}$ (total)
reflects the clockwise precession. Initially, the precession is differential, 
i.e. only the innermost parts of the torus tilt and twist. Later, during the time 
evolution, also the outer parts of the flow tilt, and the last two time snapshots 
 in the left panel of the Figure \ref{fig:tilt_5x3} show that the tilt increases 
uniformly for all radii. In the right panel of the same Figure, the slopes of the 
dashed and solid lines at late times are nearly the same, meaning that after the period of a differential precession, the torus is precessing as a nearly solid body.

 \begin{figure}
\plottwo{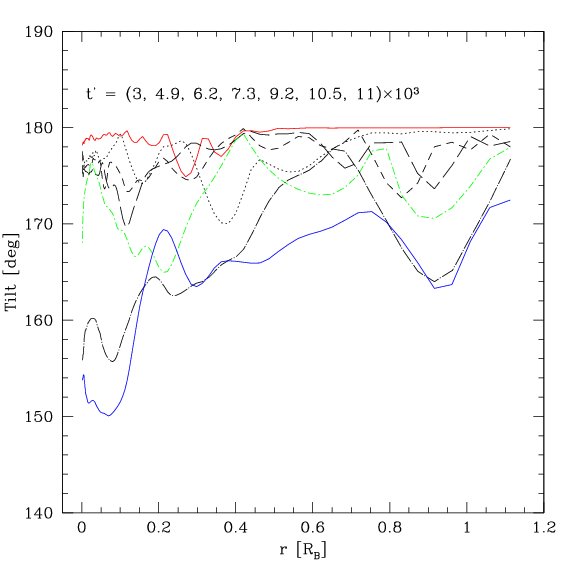}{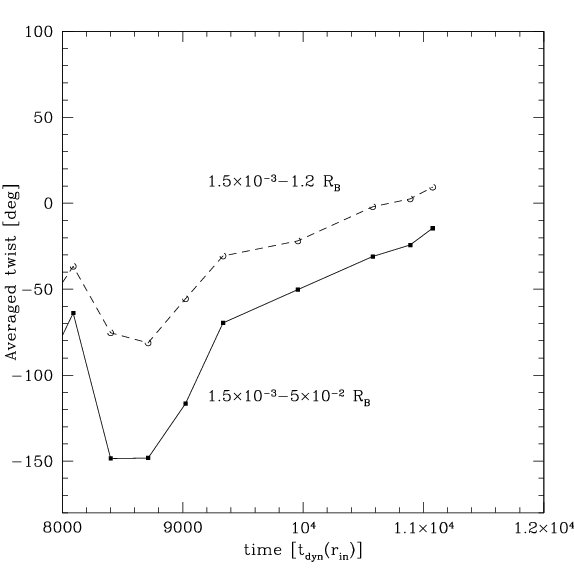}
\caption{The tilt (left panel) and averaged twist (right panel) angles,
calculated for the model
with  $\gamma=1.2$. The tilt is shown as a function of radius,
for several time-snapshots, given in the top of the panel. The initial tilt is of 
180$^{\circ}$ means that the angular momentum vector is not tilted to the z-axis 
and the 
flow rotates with a negative azimuthal velocity.
The twist angle is shown as a function of time, and the two curves 
represent the averaging over radius from $r=r_{\rm in}$ to $r=0.05 R_{\rm B}$ 
(solid line) and to $r=1.2 R_{\rm B}$ (dashed line).
The twist angle increasing from $\sim -150^{\circ}$
to  $\sim -10^{\circ}$ (innermost torus) and from
$\sim -80^{\circ}$ to  $\sim 10^{\circ}$ (total)
reflects the counter-clockwise, initially differential precession.
}
  \label{fig:tilt_1x2}
   \end{figure}

In Figure  \ref{fig:tilt_4x3} we plot the tilt and twist angles for the model with 
$\gamma=4/3$.
Here again, the  initial tilt is of 
180$^{\circ}$, meaning no tilt and negative azimuthal velocity.
The twist angle increasing from $\sim -120^{\circ}$
to  $\sim -40^{\circ}$ (innermost torus) as well as from
$\sim -70^{\circ}$ to  $\sim 0^{\circ}$ (total)
shows the counter-clockwise, initially differential, and at late times a nearly 
solid body precession.

In Figure  \ref{fig:tilt_1x2} we plot the tilt and twist evolution for $\gamma=1.2$.
The precession starts even later than in case of $\gamma=4/3$, and 
proceeds also counter-clockwise.

For the model with $\gamma=1.01$, we cannot confirm neither tilt no precession of the
torus. In Figure \ref{fig:tilt_1x01}, there is no trend of increase 
for the tilt and twist angles, and 
the large scatter in the two angles is 
caused by the virtually very small components of the angular momentum vector, 
$L_{\rm x}$ and $L_{\rm y}$. 
 We find that the tilt fluctuates periodically and no 
clear signs of precession towards any direction can be detected.

We estimated the times when precession started, and periods of precession, 
for models B,C and D. The start times are: $t^{'}=4\times 10^{3}$, $8.5\times 10^{3}$ and $9.5\times 10^{3}$, for $\gamma=5/3, 4/3$ and 1.2, respectively. The delays 
in the torus tilt are corresponding to the similar delays in the 
torus formation, in case of smaller $\gamma$ with respect to $\gamma=5/3$.
On the other hand, the precession period can be estimated at
$2.4\times 10^{4}$, $1.6\times 10^{4}$ and $6.4\times 10^{3}$.
Therefore we conclude, that the larger the adiabatic index, 
the earlier the torus forms and starts precessing, while the longer the 
precession period.

For $\gamma=1.01$ no such estimates could be made, because the torus 
shrank too much before it could start precessing.
However, we note that on the density map (Fig. \ref{fig:Maps_1x01}, middle right panel)
there is a non-axisymmetric structure, which develops in the innermost region. This
 might be a hint for an azimuthal ($m=2$) instablity mode that starts developing at 
this time in the simulations.
Therefore we made an additional test calculation, model $Av$, in which we changed the 
outer boundary condition for the velocity 
and we continuously added the angular momentum at the 
outer radius. In this way, the torus could be supported rotationally for a very long 
time during the simulation, and did not shrink like in model $A$.

\begin{figure}
\plotone{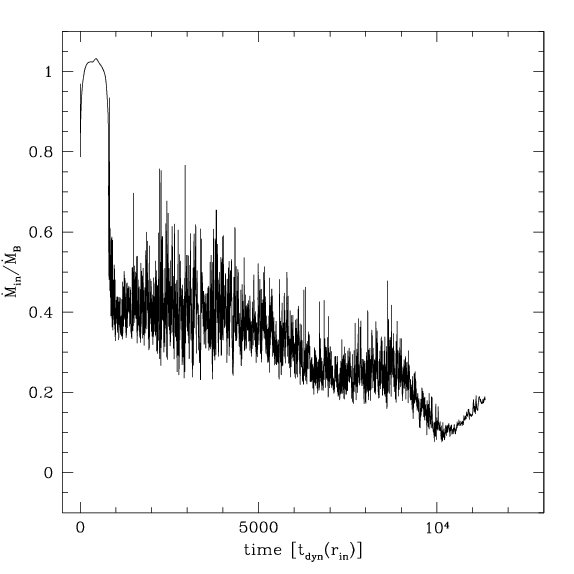}
\caption{Time evolution of the mass accretion rate
  through the inner
  boundary ($\dot M_{\rm in}$), in units of the
  Bondi rate ($\dot M_{\rm B}$). The model is $Av$, i.e. $\gamma=1.01$ 
but with the changed outer boundary condition.
}
  \label{fig:mdot_av}
\end{figure}

The accretion rate dependence on time for this model is shown in Figure
\ref{fig:mdot_av}. There is clearly no rapid rise of the accretion rate above the 
Bondi value, and the mean accretion rate is rather small, $0.3 \dot M_{\rm B}$. 
The rapid and large fluctuations of the accretion rate appear however, and temporarily
 the accretion rate flare can reach even 0.75 of the Bondi rate.
These changes are are caused by the oscillations at the
outer edge of the torus, similarly to those found in Mo\'scibrodzka \& Proga (2008)
in their 2-dimensional computations, however in our 3D case the amplitude 
of these oscillations is much larger.

\begin{figure}
\plottwo{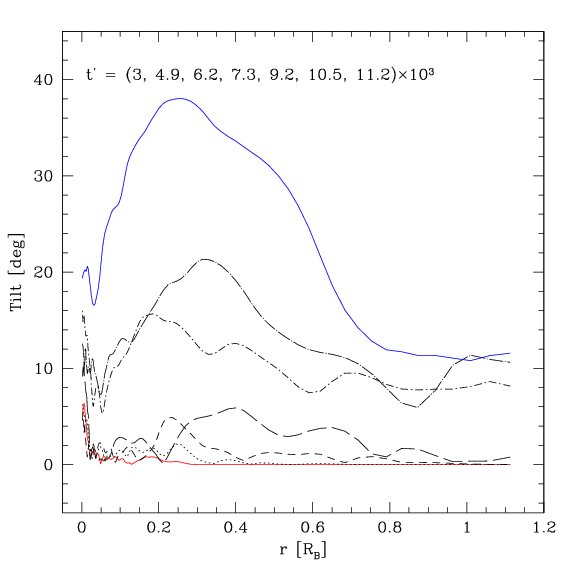}{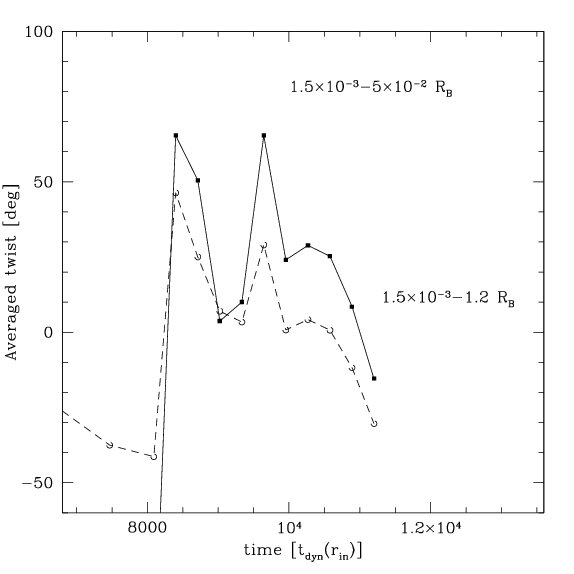}
\caption{The tilt (left panel) and averaged twist (right panel) angles,
calculated from the time evolution of the model $Av$
($\gamma=1.01$, but with the changed outer boundary condition). 
The tilt is shown as a function of radius,
for several time-snapshots, given in the top of the panel. The initial tilt of 
0$^{\circ}$ means that the angular momentum vector is not tilted and the 
flow rotates with a positive azimuthal velocity.
The twist angle is shown as a function of time, and the two curves 
represent the averaging over radius from $r=r_{\rm in}$ to $r=0.05 R_{\rm B}$ 
(solid line) and to $r=1.2 R_{\rm B}$ (dashed line).
The twist angle decreasing from $\sim 20^{\circ}$
to  $\sim -20^{\circ}$ (innermost torus) and from
$\sim 0^{\circ}$ to  $\sim -30^{\circ}$ (total)
reflects the clockwise, initially differential precession, starting from time $t' > 10^{4}$.
}
  \label{fig:tilt_1x01_av}
   \end{figure}

In Figure \ref{fig:tilt_1x01_av} we plot the tilt and twist angles for the model $Av$.
 The initial tilt is 
0$^{\circ}$, because the 
flow rotated initially with a positive azimuthal velocity. Than, the tilt increased, 
mostly in the innermost radii, and in the end of our simulation it reached $40^{\circ}$
at 0.2 $R_{\rm B}$. The tilt is initially differential, and smoothly decreases with radius.
The tilt angle for the innermost part, i.e. averaged from 
$r=r_{\rm in}$ to $r=0.05 R_{\rm B}$, initially was periodically changing. Despite 
large values of the tilt, it was due to the very small components of $L_{\rm x}$ and $L_{\rm y}$.
After time $t' > 10^{4}$, the torus started precession, and the twist angle decreased 
from $\sim 20^{\circ}$
to  $\sim -20^{\circ}$ (innermost torus) and from
$\sim 0^{\circ}$ to  $\sim -30^{\circ}$ (total).
We estimated that the precession period is below $\sim 6.3\times 10^{3}$, 
and the precession started from $t'=1.05\times 10^{4}$.
Therefore the trend of the tilt moment decreasing and the precession period increasing 
with $\gamma$ is confirmed also for $\gamma=1.01$, if the torus is supported by the 
constant inflow of the rotating material from the outer boundary.

\subsection{Evolution of the flow for various gas temperatures}
\label{sec:rbrs}

We calculated the models with various positions of the Bondi radius with respect to the 
Schwarzschild radius, which is reflected by a different sound speed at infinity and gas temperature.
In the models with smaller $R_{\rm B}/R_{\rm S}$, the evolution proceeds faster 
and the torus 
forms much earlier: at $t^{'}=4 \times 10^{2}$ for $R_{\rm B}/R_{\rm S}=300$ and at
 $t^{'}=2 \times 10^{2}$ for $R_{\rm B}/R_{\rm S}=100$.

\begin{figure}
\plotone{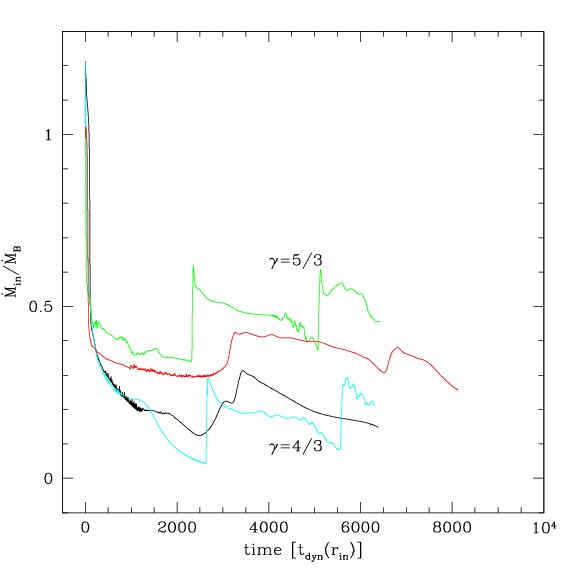}
\caption{Time evolution of the mass accretion rate
  through the inner
  boundary ($\dot M_{\rm in}$), in units of the
  Bondi rate ($\dot M_{\rm B}$). The models are: 
$E$ (red line),
$F$ (green line), $G$ (black line) and
$H$ (cyan line). The models have the adiabatic index of $\gamma=5/3$
and $\gamma=4/3$, as labelled on the plot, 
and temperature determined by the ratio of $R'=R_{\rm B}=R_{\rm S}$
is equal to 100 or 300.
}
  \label{fig:mdot_rb}
\end{figure}

During the torus evolution, in these models we detect neither tilt nor precession.
The accretion rate onto the black hole is much below the Bondi rate and it 
is fluctuating
 (Figure \ref{fig:mdot_rb}). These oscillations are caused by a strong 
outflow of material that occurs periodically because the gas wants to remove some 
angular momentum (note that the $l_{0}$ parameter, i.e. the circularisation 
radius, is the same in all models).

\begin{figure}
\epsscale{0.9}
\plotone{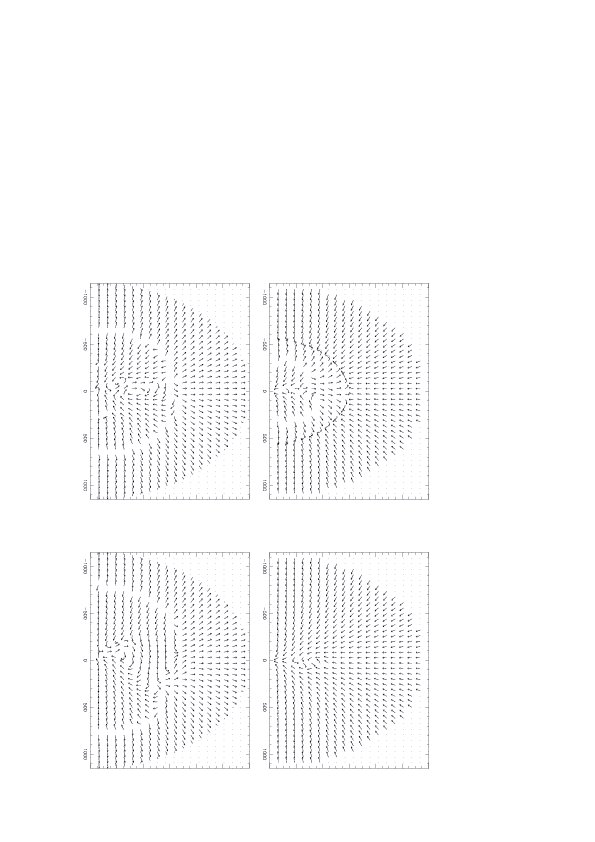}
\caption{The poloidal velocity field for the $\phi=0$ slice, 
at the end of the simulation. The models are: 
$E$ (top left),
$F$ (top right), $G$ (bottom left) and
$H$ (bottom right).
}
  \label{fig:velfield_rb100}
\end{figure}

Nevertheless, the outflows occur symmetrically in these models, and in the end of 
the simulation there are no systematic 
asymmetries with respect to the equatorial plane nor 
the azimuth.
Therefore for the models with large sound speed at infinity, the outflow stabilizes the system with respect to the azimuthal perturbations and 
there is no precession. 
The exemplary maps of the velocity field 
at the end of the simulation for the models $E$, $F$, $G$ and $H$, are shown
in Figure \ref{fig:velfield_rb100}.

\section{Discussion}
\label{sec:diss}

In this work, we considered the hydrodynamical models of 
slightly rotating, non-axisymmetric accretion flows.
We studied the role of adiabatic indices, and 
we verified the results from our previous work considering the 
precession of the inner 
torus.

We found that the tilt and precession occur not only for the gas pressure dominated 
flow, described with the equation of state with 
$\gamma=5/3$, but also for the radiation dominated gas, 
modeled with $\gamma=4/3$, as wll as for smaller $\gamma=1.2$.
The moment, when the torus gets tilted and starts precessing, correlates with $\gamma$.
The period of precession also depends on the adiabatic index, and is shorter
 for smaller $\gamma$.

We tested also the nearly isothermal gas, with $\gamma=1.01$. In this case,
we could not confirm whether the torus precesses, in the model when there is no further
angular momentum supply from the outer boundary, except for the initial condition. 
For small $\gamma$, the torus is very dense and compact,
and its evolution is determined by the large scale behaviour of the flow.
If the angular momentum is added during the simulation to the outer boundary 
of the simulation domain, such a torus may survive and keep the accretion rate onto 
black hole constantly smaller than the Bondi rate.
Otherwise, there is a lot of material that quickly 
falls radially onto the center, and because the speed of sound is very small, 
most of this material falls in supersonically. The torus shrinks 
substantially, and the net accretion rate onto the black hole is again on 
the order of the Bondi rate. The accretion rate fluctuates about this value very rapidly, because the shrunken torus still provides an obstacle for the radially
infalling gas. The boundary condition that we used for most of the models in this work,
made such a scenario of the flow evolution proceed very fast for 
small $\gamma$, and much slower for larger $\gamma$.

 The net outflow of material from the equatorial plane is
not observed in the model with $\gamma=1.01$, and at large scales the flow is 
spherically symmetric (cf. Mo\'scibrodzka \& Proga 2008).
However some tilt is visible in the innermost region, it is 
twice smaller than for $\gamma=5/3$ or $\gamma=4/3$. The twist angle 
is negligibly small, as both the $L_{\rm x}$ and
$L_{\rm y}$ components of the angular momentum vector are tiny. This 
angle fluctuates periodically, showing no clear signs of precession towards 
any direction.

However, if the angular momentum is 
constantly added at the outer boundary, the torus remains large and does not shrink. 
In such a model, the tilt angle increases after
$t'>10^{4}$ (units of dynamical times at the inner radius), 
and the torus starts precessing.
Therefore, we conclude that the non-axisymmetric perturbations occur
for all types of the equation of state and only the moment when 
instability starts growing depends on $\gamma$. If for small $\gamma$ the amount of 
inflow of the rotating material is not sufficient, the torus may shrink 
before the effects of precession can be observed.

The present work is a follow-up of Janiuk et al. (2008), where
we first identified the precession of a torus, developing in the 3-D model of a 
slowly rotating accretion flow.  We found, that if a sufficient amount of angular 
momentum is added to the initially spherically accreting Bondi flow,
the instabilities rise in the innermost gas. The initial perturbations appear
due to some small asymmetries in the angular momentum 
distribution, which in the calculations 
can be of numerical origins, but to which any physical 
system is subject. The flow then enters a transient 
phase of growing acoustic oscillations,
 which manifest themselves in the fluctuating sonic surface. Subsequently, the 
transition to a progressive departure from the initial state takes place 
(i.e. the growing tilt of the inner torus and its precession about the vertical axis).

In Janiuk et al. (2008), as well as in the present work, we checked that 
the instability develops in a highly supersonic flow, whereas for small 
Mach numbers it is suppressed. We can attribute this behaviour with the 
Papaloizou \& Pringle (1985) type of instability, which for low order modes is 
driven by the Kelvin-Helmholtz mechanism. However here, the picture is more complicated
due to the supersonic nature of the 
flow, presence of outflows and shocks, as well as compressibility.
The present work was devoted to the effect of the various adiabatic index $\gamma$.

We found that the inner torus precesses
 also for values other than $\gamma=5/3$, used in our first study:
$\gamma=4/3, 1.2$, and 1.01.
However, the time for the precession to set increases with decreasing $\gamma$.
For the nearly isothermal model, $\gamma=1.01$, the time of the instability growth
was too long to set the precession, unless the outer boundary condition was changed and
more angular momentum was supplied during the time evolution.
Our results are in agreement with the general behaviour of the unstable accreting flows
for various equations of state. Blondin et al. (2003) studied the stability 
of standing accretion shocks, and also found that the softer the equation of 
state (smaller $\gamma$), the slower is the growth of SAS instability. For the smallest
$\gamma=1.25$ used by these authors, they
 found that the shock remained marginally stable 
for a long period of time, however also this model eventually became as unstable
as the others. The physical mechanism that is lying behind this SASI instability is
the vortical-acoustic feedback between the pressure waves and vorticities induced by the
aspherical shock. Such feedback was also studied by Foglizzo (2002) 
in the context of accreting black holes. In our simulations, shocks arise due to the 
outflow of gas in the equatorial plane, and no polar outflows are produced without 
the magnetic field (see e.g. Mo\'scibrodzka \& Proga 2009), 
but the mechanism appears to be similar.

The mechanism of precession that we present in this work appears to be an additional 
one, apart from those discussed in the litearture. Caproni et al. (2006) discuss
several different sources of the accretion disk warping and precession. It may occur 
due to the tidal forces, irradiation, magnetically driven instability, 
or the Bardeen-Petterson effect. Signatures of the disk precession are observed in both
AGN and stellar mass black hole or neutron star binaries. The observations involve
the spectral and temporal variability as well as the distortions of jet morphology. 
The period of precession can be quite short, and for the sources with black 
hole mass on the order of $10^{7}-10^{8} M_{\odot}$ the observed times are
 on the order of 1-10 years.
Our results are in good agreement with these observations of AGN.

To study the precession mechanism in more detail we need to include the 
magnetic fields in our modeling. If the Kelvin-Helmholtz mechanism 
 lies behind the torus precession instability, the results should be affected
by the presence of magnetic field component only in the streaming direction.
This is planned to be the subject of our future work.
In addition, it will be 
useful to compare the results of the present ZEUS-MP code with another
MHD code available for testing, to help clarify whether the main instability 
in our studies is of a physical or numerical nature.

\section*{Acknowledgments}
We thank Bo\.zena Czerny for helpful 
discussion. We also thank the developers of ZEUS-MP
for providing the code publicly available.
This work was supported in part by grant
N N203 380136 from the Polish Ministry of Science and
in part by the National Science Foundation through 
TeraGrid resources provided by NCSA. 
DP acknowledges support provided by the Chandra awards TM8-9004X issued
by the Chandra X-Ray Observatory Center, which is operated by
the Smithsonian Astrophysical Observatory for and on behalf of NASA under
contract NAS 8-39073.
We also thank the anonymous referee, whose comments helped us to 
improve the final version of our article.

\end{document}